\documentclass[12pt,aps,prd,reprint,sort&compress, superscriptaddress, showpacs, nofootinbib, preprintnumbers]{revtex4-1}

\usepackage{graphicx}
\usepackage{amsmath}
\usepackage{amssymb}
\usepackage{dsfont}
\usepackage{amsfonts}
\usepackage[UKenglish]{babel}
\usepackage{hyperref}
\usepackage{nicefrac}
\usepackage{verbatim}
\usepackage{bbm}
\usepackage{color}
\usepackage{multirow}
\usepackage{subcaption}
\usepackage{cleveref}

\begin{document}
\title{No deconfinement in QCD ?}
 
\author{L.~Ya.~Glozman}
\email{leonid.glozman@uni-graz.at}
\affiliation{Institut f\"ur Physik, FB Theoretische Physik, Universit\"at Graz, Universit\"atsplatz 5,
8010 Graz, Austria}

\begin{abstract}
{At a critical temperature QCD in the chiral limit undergoes
a chiral restoration phase transition. Above the phase
transition the quark condensate vanishes. The Banks-Casher relation
connects the quark condensate to a density of the near-zero modes
of the Dirac operator. In the Nambu-Goldstone mode the quasi-zero
modes condense around zero, $\lambda \rightarrow 0$, and provide a nonvanishing
quark condensate. The chiral restoration phase transition implies
that above the critical temperature there is no any longer a condensation
of the Dirac modes around zero. If a $U(1)_A$ symmetry is also restored and a gap opens in the Dirac spectrum then the Euclidean correlation functions are 
$SU(2N_f) \supset SU(N_f)_L \times SU(N_f)_R \times U(1)_A$- symmetric.
This symmetry implies that a free (deconfined)
propagation of quarks in Minkowski space-time that perturbatively interact
with unconfined gluons is impossible.  This means that   QCD above the
critical temperature is not of a
quark-gluon plasma origin and has a more complicated structure.
} 
\end{abstract}
\maketitle

\section{Introduction}

Classically QCD with $N_F$ degenerate flavors in a finite volume $V$ and without exact zero modes
of the Dirac operator (which are irrelevant in the $V \rightarrow \infty$
limit) has a
$SU(2N_f) \supset SU(N_f)_L \times SU(N_f)_R \times U(1)_A$ symmetry \cite{Gloz2015}.
A symmetry of hadrons in this case is
$SU(2N_F) \times SU(2N_F)$ for mesons and 
$SU(2N_F) \times SU(2N_F) \times SU(2N_F)$ for baryons, which is a  model-independent,
pure analytical statement. The axial anomaly breaks the $U(1)_A$
symmetry, and consequently also the $SU(2N_F)$ symmetry to
$SU(N_f)_L \times SU(N_f)_R$. 
In the thermodynamic limit $V \rightarrow \infty$
the lowest lying eigenmodes of the Dirac operator condense around zero
(the so-called near-zero modes) and provide according to the Banks-Casher
relation a nonvanishing quark condensate. If effects of anomaly
and of spontaneous breaking of chiral symmetry are encoded in the same
near-zero modes, then a truncation of the near-zero modes should lead
to a large symmetry of hadrons mentioned above. 
It explains naturally previous lattice observations of emergence of a 
symmetry of hadrons, that is larger than  the chiral $SU(N_f)_L \times SU(N_f)_R \times U(1)_A$
symmetry of the QCD Lagrangian, after an artificial subtraction of the near-zero
modes of the Dirac operator 
\cite{Denissenya:2014poa,Denissenya:2014ywa,Denissenya:2015mqa,Denissenya:2015woa}. 

These results have a very nontrivial implication for QCD above the chiral
restoration phase transition at a critical temperature $T_c$. Here we discuss
the chiral limit, where QCD undergoes a chiral restoration phase transition
\cite{PW} (at finite quark masses the phase transition converts into a fast
cross-over, as it follows from  lattice measurements).

In this note we address QCD at zero chemical potential, because in this case
there exists a rigorous connection between the quark condensate of the vacuum
and a density of modes of the Dirac operator around zero -
 the
Banks-Casher relation.  In the Nambu-Goldstone mode of
chiral symmetry, i.e. below the phase transition, the modes of the Dirac operator condense
around zero, $\lambda \rightarrow 0$, and provide a nonvanishing quark condensate.

Above the phase transition the quark condensate of the vacuum vanishes. 
If in addition the $U(1)_A$ symmetry is also restored and
 a gap opens in the Dirac spectrum around  $\lambda = 0$ \cite{A1,A2}\footnote{See also the opposite statement \cite{K} and its
 critique in ref. \cite{A2}. We will assume that conclusions of 
 the JLQCD collaboration \cite{A1,A2} are correct \cite{Brandt:2016daq}.}
 then the Euclidean  QCD correlation functions become  $SU(2N_f)$
symmetric. Such a symmetry prohibits in Minkowski space-time a propagation of
a free deconfined massless quark that perturbatively interact with gluons, because the magnetic interaction manifestly  breaks
 the $SU(2N_f)$ symmetry. The only
possibility is that the chirally symmetric quarks are confined  
into $SU(2N_F) \times SU(2N_F)$ and
$SU(2N_F) \times SU(2N_F) \times SU(2N_F)$ symmetric "hadrons".

\section{$SU(2N_F)$ hidden classical symmetry of  QCD}

In this section we review some results of ref. \cite{Gloz2015}.
Nonperturbatively QCD is defined in Euclidean space-time in a finite box
with a volume $V$  with the lattice
ultraviolet regularization.
Consider  the Lagrangian in Euclidean space-time for $N_F$ degenerate quark
 flavors in a given gauge background,

\begin{equation}
{\cal {L}} = \Psi^\dag(x)( \gamma_\mu D_\mu + m) \Psi(x)
\label{lag}
\end{equation}

\noindent
with

\begin{equation}
D_\mu = \partial_\mu + i g\frac{t^a}{2} A^a_\mu,
\end{equation}

\noindent
where $A^a_\mu$ is the gluon field configuration and
$t^a$ are the $SU(3)$-color generators.  
Note that in Euclidean space-time only the Hermitian conjugation, $\Psi^\dag(x)$,
can be used (not a Dirac adjoint bispinor) and the fields  $\Psi^\dag(x)$ and  $\Psi(x)$ are independent
from each other.\footnote{Very often instead of $\Psi^\dag$ the $\bar \Psi$
notation is used in Euclidean space. Then it should be kept in mind that under
Euclidean Lorentz transformations ($SO(4)$) $\bar \Psi$ transforms as $\Psi^\dag$.}

The hermitian Dirac operator for a quark in a given gluonic configuration, $i \gamma_\mu D_\mu$, 
has in a finite volume a discrete spectrum with real eigenvalues $\lambda_n$:

\begin{equation}
i \gamma_\mu D_\mu  \Psi_n(x) = \lambda_n \Psi_n(x).
\label{ev}
\end{equation}

\noindent
The nonzero eigenvalues come in pairs $\pm \lambda_n$,
because 

\begin{equation}
i \gamma_\mu D_\mu  \gamma_5 \Psi_n(x) = -\lambda_n \gamma_5 \Psi_n(x).
\end{equation}

\noindent
We  define  independent fields $\Psi(x)$ and $\Psi^\dagger(x)$ in (\ref{lag}) in the following standard way. Namely we expand them over a complete
and orthonormal set $\Psi_n(x)$:

\begin{equation}
 \Psi(x) = \sum_{n} c_n \Psi_n(x), ~~~~~ 
 \Psi^\dagger(x) = \sum_{k}\bar {c}_k \Psi^\dag_k(x),
\end{equation}
 where $\bar {c}_k,c_n$ are independent Grassmann numbers.
Then the fermionic part of the QCD partition function  takes the following form

\begin{equation}
Z = \int \prod_{k,n} d\bar {c}_k dc_n  
e^{\sum_{k,n}\int d^4x 
 \bar {c}_k  c_n (\lambda_n + im) \Psi_k^\dag(x) \Psi_n(x)}. 
\label{ZZ}
\end{equation}

Now we will analyse symmetry properties of the partition function
above assuming that there are no exact zero modes.
The $SU(2)_{CS} \supset U(1)_A$  ($CS$ - {\it chiralspin} )transformations are
defined  as \cite{Glozman:2014mka,Glozman:2015qva}

\begin{align}
\label{V-def}
  \Psi \rightarrow  \Psi^{'} &= e^{i  \frac {\boldsymbol{\varepsilon} \cdot \boldsymbol{\Sigma}}{2}} \Psi , \; 
\end{align}

\noindent
with the following generators 
\begin{align}
\label{sigma}
\boldsymbol{\Sigma} = \{ \gamma^4, i \gamma^5 \gamma^4, -\gamma^5 \} \; ,  
\end{align}

\noindent
that form an $SU(2)$ algebra
\begin{align}
 [\Sigma^i,\Sigma^j] = 2 i \epsilon^{i j k} \, \Sigma^k \; .
\end{align}

\noindent
  The $\gamma^4$ and $\gamma^5 \gamma^4$ matrices mix the right- and left-handed components of the fermion fields.

We can combine the $SU(2)_{CS}$ rotations with the flavor $SU(N_F)$ transformations into one larger group, 
$SU(2N_F)$. In  the case of two flavors the $SU(4)$ transformations

\begin{align}
\label{W-def}
\Psi \rightarrow  \Psi^{'} &= e^{i \boldsymbol{\epsilon} \cdot \boldsymbol{T}/2} \Psi\; ,
\end{align}
are defined through the following set of 15 generators:
\begin{align}
 \{(\tau^a \otimes \mathds{1}_D), (\mathds{1}_F \otimes \Sigma^i), (\tau^a \otimes \Sigma^i) \} \;, 
\end{align}

\noindent
where $\tau^a$ are isospin Pauli matrices.
If the $(2N_F)^2 -1$-dimensional  rotation vector
$\boldsymbol{\epsilon}$ is a constant for the whole 3+1-dim space, then the corresponding transformation
 is global, while with
the space-dependent rotation $\boldsymbol{\epsilon}( \boldsymbol{x})$ it is local. Obviously, {\it the Lagrangian (1) does not have these symmetries, because
the Dirac operator does not commute with the $SU(2)_{CS}$ generators}.

The eigenmodes of the Dirac operator in a finite volume $V$ can be separated into two classes. The nonzero eigenmodes, $\lambda_n \neq 0 ~~(n \neq 0)$, and
 exact zero modes, $\lambda =0~~ (n=0)$.
The exact zero modes satisfy  the Dirac
equation,

\begin{equation}
 \gamma_\mu D_\mu  \Psi_0(x) = 0.
\label{dir}
\end{equation}

\noindent
They are chiral, $L$ or $R$. 
Given standard antiperiodic boundary conditions for fermions
along the time direction the zero modes
appear only in gauge configurations with a nonzero global
topological charge. The difference of numbers of the left-handed and right-handed
zero modes is fixed according to the Atiyah-Singer theorem by the global topological charge of the gauge configuration. 
Consequently there is no one-to-one correspondence of the left-
and right-handed zero modes:
The zero modes induce an asymmetry between the left and the right
at nonzero global topological charge. The $SU(2)_{CS}$ chiralspin
rotations (except for pure $U(1)_A$ rotations) mix the left- and right-handed
Dirac spinors. Such a mixing can be defined only if there is a 
one-to-one mapping of the left- and right-handed Dirac spinors in the
Hilbert space.
In other words the zero modes explicitly violate the
$SU(2)_{CS}$ invariance, see Appendix for details. The zero modes are the precise reason for absence of the $SU(2)_{CS}$ invariance in the exponential
(classical) part of the partition function (6).

It is well understood, however,  that these exact zero
modes are completely irrelevant since their contributions
to the Green functions and observables vanish in the thermodynamic
limit $V \rightarrow \infty$ as $1/V$, see e.g. refs. \cite{LeuS,N,A3}. 
Consequently, in the finite-volume calculations we can ignore (or subtract) the exact zero-modes.

Now we can directly read off the symmetry properties
of the partition function (\ref{ZZ}) on the subspace that does
not include the zero modes..
For any 
$SU(2)_{CS}$ and $SU(2N_F)$ transformation  the 
$\Psi_n$ and $\Psi^\dag_m$  Dirac bispinors transform as

\begin{equation}
\Psi_n \rightarrow U \Psi_n,~~~\Psi^\dag_m \rightarrow (U \Psi_m)^\dag,
 \end{equation}

\noindent
where $U$ is any unitary transformation  from the groups
$SU(2)_{CS}$ and $SU(2N_F)$ , $U^\dagger = U^{-1}$.
It is then obvious that the exponential part of the partition function,
which is a functional with fixed $\lambda_n, c_n, \bar c_k$,
{\it is invariant with respect to global and local $SU(2)_{CS}$ 
and $SU(2N_F)$ transformations}, because

\begin{equation}
(U\Psi_k(x))^\dag U \Psi_n(x) = \Psi^\dag_k(x) \Psi_n(x).
\label{tt}
\end{equation}
 
\noindent
This equation is well defined on a subspace of Dirac modes
that does not include the zero modes, see Appendix for a proof.
The exact zero modes  contributions 
$$\Psi^\dag_0(x) \Psi_n(x),\Psi^\dag_k(x) \Psi_0(x), \Psi^\dag_0(x) \Psi_0(x),$$ for which the equation (\ref{tt}) is not
defined, are irrelevant in the thermodynamic limit and can be ignored.\footnote{The
exact zero modes break the $SU(2)_{CS}$ and $SU(2N_F)$
symmetries, see Appendix. Consequently these symmetries are absent at the QCD
Lagrangian level.}
In other words, QCD classically without the irrelevant exact zero  modes
 has  in a finite volume $V$ both global and local $SU(2N_F)$ 
 symmetries. These  $SU(2)_{CS}$ 
and $SU(2N_F)$ symmetries are hidden\footnote{Because they are not seen at the
Lagrangian level and become visible only when the irrelevant
exact zero modes are subtracted.} classical symmetries of QCD.
We emphasize, to avoid any confusion, that the symmetry properties
of the Dirac operator in (1) and the symmetry properties of the classical
part of the partition function on the subspace that does not include
the exact zero modes need not be the same. Only with the full Hilbert
space of the Dirac eigenmodes, that necessarily includes zero modes, these
$SU(2)_{CS}$ and $SU(2N_F)$
symmetries must be absent in the integrand. This has been proven in the Appendix. However, once we consider a subspace of the eigenmodes there is no
a general constraint on a symmetry.
And indeed, we have demonstrated
that a symmetry of the integrand in (6), once the zero modes are ignored,
is higher - it is $SU(2)_{CS}$ and $SU(2N_F)$ symmetric, contrary to the
Lagrangian (1).

The axial anomaly, that stems from the noninvariance of the
measure $\prod_{k,n} d\bar {c}_k dc_n$ under a local $U(1)_A$ transformation
\cite{FU},
 breaks the classical $U(1)_A$ symmetry. Since the
$U(1)_A$ is a subgroup of $SU(2)_{CS}$, the axial anomaly breaks either
the $SU(2)_{CS}$ and $SU(2N_F) \rightarrow SU(N_F)_L \times SU(N_F)_R$.
In other words, {\it the fermion determinant is not invariant under 
$SU(2)_{CS}$ and $SU(2N_F)$ because it does contain effect of the anomaly}.
 
In the thermodynamic limit $V \rightarrow \infty$ the otherwise finite
lowest eigenvalues $\lambda$ condense around zero and provide
according to the Banks-Casher relation at $m \rightarrow 0$ a nonvanishing
quark condensate in Minkowski space \cite{BC}

\begin{equation}
\lim_{m \rightarrow 0} <0|\bar \Psi(x) \Psi(x)|0> = -\pi \rho(0).
\end{equation}

\noindent
 Here a sequence of limits is important: first 
an infinite volume limit is taken
 and only then - a chiral limit.  
The quark condensate  in Minkowski space-time, $<0|\bar \Psi(x) \Psi(x)|0>$, breaks all $U(1)_A$, $SU(N_F)_L \times SU(N_F)_R$, $SU(2)_{CS}$
 and $SU(2N_F)$ symmetries to $SU(N_F)_V$.
Consequently, the hidden classical $SU(2)_{CS}$ and $SU(2N_F)$ symmetries
are broken both by the anomaly and spontaneously.

\section{Restoration of $SU(2)_{CS}$ and $SU(2N_F)$ at high temperature} 

The hidden classical $SU(2)_{CS}$ and $SU(2N_F)$ symmetries are
broken by the anomaly and by the quark condensate. Above the chiral
restoration phase transition the quark condensate vanishes. If in
addition the $U(1)_A$ symmetry is also restored \cite{A1,A2} and
a gap opens in the Dirac spectrum, then
it follows that above the critical temperture the $SU(2)_{CS}$ and $SU(2N_F)$
symmetries are manifest. The precise meaning of this statement is that the
correlation (Green) functions and observables are $SU(2)_{CS}$ and $SU(2N_F)$ 
symmetric.

\section{ No free deconfined quarks above the chiral restoration phase transition}

 What do these $SU(2)_{CS}$ and  $SU(2N_F)$ symmetries of Euclidean
 correlation functions imply
for Minkowski space-time, where we live?
They imply that there cannot be deconfined free quarks and gluons at any
finite temperature. 

Assume that above $T_c$ QCD is
in a deconfined phase. Then, according to the definition of deconfinement
and of the quark-gluon plasma phase, there must be free propagating quarks. Free propagating quarks, 
interacting with gluons, 
are solutions of the Dirac equation and have the following Lagrangian

\begin{equation}
\label{Lagrangian}
 \overline{\Psi}  i \gamma^{\mu} D_{\mu} \Psi = \overline{\Psi}  i \gamma^0 D_0  \Psi 
  + \overline{\Psi} i \gamma^k D_k  \Psi\; . 
\end{equation}

\noindent
The first term describes an interaction of the quark charge density 
 with the chromo-electric  
part of the gluonic field and the second term contains a kinetic term for a free quark as well as  an
interaction of the spatial current density with the chromo-magnetic field.

While the chromo-electric part of the Dirac Lagrangian is invariant under  global and space-local  
$SU(2)_{CS}$ and
$SU(2N_F)$ transformations, the kinetic term and the quark - chromo-magnetic field
interaction  - are not.  Consequently the Green functions and observables calculated
in terms of unconfined quarks and gluons in Minkowski space (i.e. within the perturbation theory) cannot be $SU(2)_{CS}$ and
$SU(2N_F)$ symmetric at any finite temperature, because the magnetic
interaction necessarily breaks both symmetries.

{\it Also above $T_c$ QCD is in a confining regime.}

In contrast,  color-singlet $SU(2N_F)$-symmetric "hadrons" (with not yet known properties) 
are not prohibited by the restored  hidden $SU(2N_F)$ symmetry of  QCD  and can freely propagate. "Hadrons" with such a symmetry in Minkowski 
space can be
constructed \cite{Sh}.

\section{Discussion}

Early arguments about deconfinement at high temperature and
transition to the quark-gluon plasma are all based on perturbative
derivation of a Debye screening of the color charge. QCD is however
never perturbative and what these perturbative calculations mean is
not clear. Such calculations are self-contradictory: They rely on
unconfined quarks and gluons and at the same time try to address
confining properties without any clear definition what deconfinement
would mean. While something drastic might indeed happen with gluodynamics at
high $T$, whether it means deconfinement or not is by far not clear.

The Wilson and Polyakov loop criteria of confinement-deconfinement
are applicable only for a pure glue theory. While lattice measurements
of the Wilson and Polyakov loops (and of related $Z_3$ symmetry)  do 
show that indeed  some properties of a pure glue theory rapidly change
at the critical temperature, it is by far not clear whether it means
deconfinement or not. To conclude about confinement-deconfinement one
invokes as an intermediate step an {\it interpretation} of the Wilson loop
and of a correlator of the Polyakov loops as a "potential" between the static
color charges. What this "potential" would mean for quarks that move 
and whether they are confined or not is not clear.

Here in contrast we rely  on the truly nonperturbative and rigorous Banks-Casher
relation and on a symmetry of  QCD  above the chiral restoration
phase transition. Namely, we have shown that 
given manifest $SU(N_F)_L \times SU(N_F)_R$ and $U(1)_A$ chiral symmetries
the actual symmetry of QCD with $N_F$
degenerate flavors is $SU(2N_F)$ that prohibits in Minkowski space-time
an on-shell propagation of  free deconfined quark that interact with perturbative gluons.

\section{Predictions}

Appearance of the $SU(2)_{CS}$ and of $SU(2N_F)$ symmetries at
$T > T_c$ can be directly tested on the lattice. By definition
QCD is said to be symmetric  under some symmetry group $U$ if the diagonal
correlation functions calculated with a set of operators ${O_1,O_2,...}$
that form an irreducible representation of the group $U$ are identical, and
if the off-diagonal cross-correlators vanish. 

Transformation properties
of meson and baryon operators under $SU(2)_{CS}$ and  $SU(2N_F)$ groups are given
in refs. \cite{Glozman:2015qva,Denissenya:2015woa}. 
In particular, three isovector $J=1$ mesonic operators
$\bar \Psi \vec \tau \gamma^i \Psi,(1^{--})$; 
$\bar \Psi \vec \tau \gamma^0 \gamma^i \Psi,(1^{--})$;
$\bar \Psi \vec \tau \gamma^0 \gamma^5 \gamma^i \Psi,(1^{+-})$
form an irreducible representation of $SU(2)_{CS}$. One expects that
below $T_c$ all three diagonal correlators will be different and
the off-diagonal cross-correlator of two $(1^{--})$ operators will
be not zero. Above $T_c$ a $SU(2)_{CS}$ restoration requires that
all three diagonal correlators should become identical after a common normalization
at some point  and the off-diagonal correlator of
two $(1^{--})$ currents should vanish. 
A restoration of $SU(2)_{CS}$ and of $SU(N)_L \times SU(N)_R$ 
(the latter can be tested e.g. through a coincidence of the diagonal correlators
with the vector- and axial-vector currents)
implies a restoration of $SU(2N_F)$.

A similar prediction can be made
with the baryon operators.

\begin{acknowledgements}

The author thanks T. D. Cohen, C. Gattringer and C. B. Lang for a careful
reading of the ms.
We acknowledge partial support from the Austrian Science Fund (FWF)
through the grant P26627-N27. 
\end{acknowledgements}

\section{Appendix}

In this appendix we prove the following statements:

(i) The $SU(2)_{CS}$ and $SU(2N_F)$ transformations are well
defined on the subspace of the nonzero eigenmodes of the Dirac
operator and this subspace is invariant under these transformations.

(ii)These transformations are not defined on the full Hilbert
space of Dirac eigenmodes that includes the zero modes.

For convenience we will use the chiral representation  and 
generators of $SU(2)_{CS}$ are the following matrices:

\begin{align}
 \gamma_4 = \begin{pmatrix}
  0 & 1 \\
  1 & 0
 \end{pmatrix} \; ,
\end{align}

\begin{align}
 \gamma_5 \gamma_4 = \begin{pmatrix}
  0 & -1 \\
  1 & 0
 \end{pmatrix} \; ,
\end{align}

\begin{align}
 \gamma_5 = \begin{pmatrix}
  -1 & 0 \\
  0 & 1
 \end{pmatrix} \; .
\end{align}

The left- and right-handed projections are defined as

\begin{equation}
\Psi_L = \frac{1-\gamma_5}{2} \Psi; ~~~~\Psi_R = \frac{1+\gamma_5}{2} \Psi
\end{equation}

\noindent
with

\begin{align}
\Psi= \begin{pmatrix}
  \chi_L \\
  \chi_R   
 \end{pmatrix} \; .
\end{align}

It is clear that the $\gamma_4$ and $\gamma_5 \gamma_4$
generators of $SU(2)_{CS}$ mix the left- and right-handed spinors:

\begin{equation}
\gamma_5\Psi_L = - \Psi_L; ~~~~ \gamma_5\Psi_R = \Psi_R;
\end{equation}

\begin{equation}
\gamma_4\Psi_L =  \Psi_R; ~~~~ \gamma_4\Psi_R = \Psi_L;
\end{equation}

\begin{equation}
\gamma_5 \gamma_4\Psi_L =  \Psi_R; ~~~~ \gamma_5 \gamma_4\Psi_R = -\Psi_L.
\end{equation}

\noindent
Then, given (3) and (4) for all $n \neq 0$ we have:

\begin{equation}
{\Psi_n}_L = \frac{1}{2}\Psi_n - \frac{1}{2}\Psi_{-n},
\end{equation}

\begin{equation}
{\Psi_n}_R = \frac{1}{2}\Psi_n + \frac{1}{2}\Psi_{-n},
\end{equation}

\noindent
where $\Psi_{-n} = \gamma_5 \Psi_n$.
Consequently, on the subspace $n \neq 0$ of the full Hilbert space
there is a one-to-one mapping of the left- and right-handed spinors
and in this case the $SU(2)_{CS}$ transformations are well defined.
It is also obvious that this subspace is invariant under $SU(2)_{CS}$
and $SU(2N_F)$ because any element from these groups transforms a vector
from this subspace to one and only one vector of the same subspace.
This concludes the proof of the statement (i).

 For a noninteracting
fermionic field a general form of
$\Psi_n$ at $n \neq 0$ is

\begin{align}
\Psi_n= \begin{pmatrix}
  \chi \\
  \chi   
 \end{pmatrix} \; ,
\end{align}
where $\chi$ is a two-component spinor.

Now we will prove the statement (ii). We will consider for simplicity
only the $Q=1$ sector ($Q$ is the global topological charge of the
gauge configuration), that contains one left-handed zero mode and
does not contain any right-handed zero mode. A generalization to any
$Q \neq 0$ is obvious.

The zero mode solution in this case is

\begin{equation}
\Psi_0 \equiv {\Psi_0}_L.
\end{equation}

\noindent
If we act with the $SU(2)_{CS}$ generators on this spinor
we obtain:

\begin{equation}
\gamma_5 {\Psi_0}_L = -{\Psi_0}_L,~~ \gamma_4 {\Psi_0}_L = \phi,~~
\gamma_5 \gamma_4 {\Psi_0}_L = \phi,
 \end{equation}

\noindent
where $\phi$ is some  right-handed spinor that does not
satisfy the Dirac equation, $\gamma_\mu D_\mu \phi \neq 0$,
because there is no right-handed zero mode in the $Q=1$ sector.

Now we will prove that $\phi$ does not belong to the Hilbert space.
Assume that it does. Then it can be expanded over the right-handed
non-zero modes:

\begin{equation}
\phi =\sum_{n=\pm 1, \pm 2,...} \alpha_n {\Psi_n}_R.
\end{equation}

\noindent
We multiply this equation with $\gamma_4$:

\begin{equation}
\gamma_4 \phi =\sum_{n=\pm 1, \pm 2,...} \alpha_n \gamma_4 {\Psi_n}_R
\end{equation}

\noindent
or 
\begin{equation}
 {\Psi_0}_L =\sum_{n=\pm 1, \pm 2,...} \alpha_n  {\Psi_n}_L.
\end{equation}

\noindent
The latter relation means that ${\Psi_0}_L$ should be linear dependent
with a set ${\Psi_n}_L,~~n \neq 0$. The latter is however not
true since the zero and nonzero modes form a linear independent
basis of the Hilbert space. Consequently, our assumption is not true.
In other words, $\phi$ is a spinor that is outside the Hilbert space.
This means that the $SU(2)_{CS}$ and $SU(2N_F)$ transformations
are not defined on the full Hilbert space that includes the zero modes.

\noindent
Q.E.D.

\end{document}